\documentclass[twocolumn,pra,showpacs,preprintnumbers,amsmath,amssymb]{revtex4}

\usepackage{graphicx}
\usepackage{amsmath, amsfonts, amssymb, amsthm} 
\usepackage{graphicx, psfrag}			
\usepackage[font=small]{caption}
\usepackage[captionskip=3pt,farskip=10pt,justification=justified,font=small]{subfig}
\usepackage[sort&compress]{natbib}

\newcommand{\vc}[1]{\mathbf{#1}}

\newcommand{\RR}{\mathbb{R}}

\newcommand{\Raines}{R09}

\begin{document}


\title{Feasibility of 3D reconstructions from a single 2D diffraction measurement}

\author{Pierre Thibault}
 \email{pierre.thibault@ph.tum.de}
\affiliation{%
Physics department, Technical University of Munich, Garching, Germany 
}%
\altaffiliation{%
Previous address: Paul Scherrer Institut, 5232 Villigen PSI, Switzerland
}%

\date{\today}

\begin{abstract}
We comment on the recent manuscript by Raines et al. [arXiv:0905.0269v2] (now published in Nature, vol. 463, p. 214-217, 2010), which suggests that in certain conditions a single diffraction measurement may be sufficient to reconstruct the full three-dimensional density of a scatterer. We show that past literature contains the tools to assess rigorously the feasibility of this approach. We question the formulation of the reconstruction algorithm used by the authors and we argue that the experimental data used as a demonstration is not suitable for this method, and thus that the reconstruction is not valid.

This second version was produced for documentation purposes. In addition to the minimally modified original comment, it includes in appendix a subsequent reply to one of the authors (J. Miao).
\end{abstract}



\maketitle

\section{Introduction}

This short article is a comment on the work recently reported by Raines, Salha, Sandberg, Jiang, Rodr\'iguez, Kapteyn, Du and Miao
\cite{Raines2009} (from now on referred to as \Raines) on the extension of diffraction microscopy to three-dimensional densities based on a single large-scattering-angle measurement. 
The authors claim that the approach that they called ``ankylography'' allows the full reconstruction of the three-dimensional density of a compact object from a single highly oversampled diffraction measurement, thus overcoming the need of combining many diffraction datasets, as is typically done in tomography. 

This report aims at promoting further reflection on the subject. The first section covers in some details what we believe is a much firmer basis to assess the feasibility of data extension from an Ewald sphere sampling, and introduces some of the past literature that was unfortunately overlooked by \Raines. In particular, early work on \emph{non-uniform sampling} provide valuable analytical tools for this problem. We also point to the close relationship between the current far-field approach and ``digital inline holography'', which can also produce three-dimensional information from two-dimensional measurements. Our paper also addresses what we believe are two important shortcomings of the manuscript: the reconstruction algorithm and the experimental demonstration of the method. 




\section{A sampling problem}

\subsection{Preliminaries}

In what follows, $\rho$ represents the density of an isolated specimen and is element of $L^2(\RR^D)$, with $D = 1,2, \text{or } 3$ and has compact support $\mathcal{S}$.

The Fourier transform operation, denoted $\mathcal{F}$, will be defined as 
\begin{equation}
\tilde f (q) = \mathcal{F} f(x) = \frac{1}{\sqrt{2 \pi}} \int_{\RR} f \, e^{-ixq} \, dx,
\label{eq:FT1D} 
\end{equation}
with a straightforward generalization to higher dimensions $D > 1$.

The autocorrelation of $\rho$ is defined as 
\begin{equation}
A(\vc{r}) = \rho \ast \rho = \mathcal{F}^{-1} \left| \tilde \rho \right|^2
\end{equation}
and has support $\mathcal{S}_{A} = \mathcal{S} - \mathcal{S}$. The support of the autocorrelation extends along all axes over twice the range of $S$. For instance, in 3D if $\mathcal{S}$ is included in a box of dimensions $w_x \times w_y \times w_z$, then $\mathcal{S}_{A}$ is contained in a box of dimensions $2w_x \times 2w_y \times 2w_z$. 

The objective of diffraction microscopy is to reconstruct $\rho(\vc{r})$ given the square of its Fourier transform, $\left| \tilde \rho \right|^2$. Because $A$ has compact support, $\left| \tilde \rho \right|^2$ is a band-limited function. In most instances, $\left| \tilde \rho \right|^2$ is assumed to be completely known up to a cutoff resolution, that is, it is sampled on a grid with a spacing $\Delta q_x$ finer than the Nyquist interval $\Delta q_x < \pi/w_x$ (with the same relations for the second and third dimensions). In this context, the density $\rho$ recovered by phase retrieval techniques was shown \cite{Bruck1979,Bates1982} to be unique with a probability $1$ in dimensions $D \geq 2$. 

\subsection{Scattering in the Born approximation}

We restrict the discussion to $D=3$. Let the incident planar wavefront with wavenumber $k$ propagate along the $z$ axis. The two-dimensional manifold defined by
\begin{equation}
 \left| \vc{q} - k \hat{\vc{z}} \right|^2 = k^2
\end{equation}
is called the \emph{Ewald sphere}. A central result of the kinematic theory of scattering relates the two-dimensional distribution of scattered intensities to the three-dimensional Fourier transform of $\rho$:
\begin{equation}
 I(\vc{q}_{\perp}) \propto \left| \tilde \rho (\vc{q}) \right|^2 \Big|_{\text{Ewald}},
\label{eq:IEwald}
\end{equation}
where $\vc{q} = (\vc{q}_{\perp}, q_z)$ is the Fourier space coordinate divided in its perpendicular and parallel components, and the label \emph{Ewald} indicates that the function is evaluated on the Ewald sphere (neglecting backward scattering):
\begin{equation}
 f(\vc{q}) \Big|_{\text{Ewald}} = f(\vc{q}_{\perp}, q_z = k - \sqrt{k^2 - \vc{q}^2_{\perp}}).
\end{equation}
Because $I$ provides a partial measurement of $\left| \tilde \rho  \right|^2$, the full three-dimensional density can be approximated with multiple measurements at different incident angles. This is the approach used in crystallography, as well as the prevailing technique in diffraction microscopy \cite{Miao1998,Miao2002a,Chapman2006,Barty2008}.

\Raines~suggest that a single intensity measurement may suffice to reconstruct $\rho$, even though $I$ provides only a \emph{highly non-uniform} sampling of $\left| \tilde \rho  \right|^2$.

The proposed method is similar in some aspects to another technique called digital inline holography \cite{Xu2001,Rosenhahn2009}, which consists of extracting three-dimensional information from a single near-field measurement (an inline hologram). While the measurement conditions are different, the depth information is fundamentally encoded in the same way since it is the non-negligible curvature of the Ewald sphere that gives a focal depth that is much smaller than the depth of the object. Computed propagation of the inline hologram allows determining the position of scatterers along the propagation direction. Yet, what can be obtained from digital inline holography is rarely more than ``shadows'', which are sufficient for particle tracking applications, for instance, but are not suitable for full 3D density reconstructions (see \textit{e.g.} \cite{Pu2005}).

\subsection{Non-uniform sampling}

In terms of the three-dimensional autocorrelation function, equation \eqref{eq:IEwald} is
\begin{equation}
 I = \mathcal{F} A \Big|_{\text{Ewald}},
\label{eq:IEwald2}
\end{equation}
where the function arguments were omitted. We now consider how $\left| \tilde \rho  \right|^2$ can be extended from the single measurement $I$, without additional assumptions such as the reality or the positivity of $A$ and $\rho$. In principle, continuation from the Ewald sphere to $\RR^3$ is always possible, since $I$ is an analytic function. In practice however, extension will fail unless very strong requirements are satisfied.

A large body of literature addresses the problem of interpolating band-limited functions, with its most famous probably being Shannon's 1949 work \cite{Shannon1949}. While Shannon initially considered only equally spaced sampling, many have since worked on the more difficult question of non-uniform sampling. Early work by Duffin and Shaeffer \cite{Duffin1952} on the subject are still being used, especially in the field of ``super-resolution'', where multiple images are combined to increase resolution. 
Yen \cite{Yen1956} provided exact interpolating formula for important special cases. In 1967, Landau \cite{Landau1967} showed how the Nyquist frequency criterion can be generalized to non-uniform sampling.

While general results are obtained in $L^2$ space, much of the recent work on non-uniform sampling for applications uses a finite-dimensional version of the linear equation \eqref{eq:IEwald2}, where the Fourier transform becomes a Fourier series \cite{Bucci1993, Uzunov2006} (or ``trigonometric polynomial'' \cite{Strohmer2000}). In this approach, the autocorrelation is sampled on a rectangular grid and $\left| \tilde \rho  \right|^2$ is made periodic, an approximation valid only if its magnitude decays sufficiently on the edge of the repeating box.

Let $\{ \vc{r}_1, \vc{r}_2, \dots, \vc{r}_{N} \}$ be the finite sequence resulting from an arbitrary labeling of the grid points belonging to $\mathcal{S}_A$. Let in addition $\{\vc{q}_1, \vc{q}_2, \dots, \vc{q}_{M} \}$ be the sequence of Fourier space coordinates of the measured intensity samples, again with an arbitrary labeling. Define two vectors $y \in \RR^M$ and $x \in \RR^N$ such that $y_i = I(\vc{q}_i)$ and $x_i = A(\vc{r}_i)$, and let the Fourier transform operation become a $N \times M$ matrix 
\begin{equation}
F_{ij} = \frac{1}{\sqrt{N}} \exp \left( -i \vc{q}_i \cdot \vc{r}_j \right).
\end{equation}
Equation \eqref{eq:IEwald2} is now cast as a finite-dimensional linear system:
\begin{equation}
y = F x.
\label{eq:xfy}
\end{equation}
The problem of recovering the three-dimensional autocorrelation from the intensity measurement is ill-posed if $N > M$. In principle, if $M > N$ and $F$ is full-rank, $x$ can be retrieved exactly. Yet, the linear system probably has no solution because of the uncertainty in the measurement in $I$. The least-square solution is given by
\begin{equation}
x_{\text{ls}} = F^{+} y,
\label{eq:Finv}
\end{equation}
where $F^{+} = \left(F^{\dagger} F \right)^{-1} F^{\dagger}$ is the Moore-Penrose pseudoinverse of $F$. That $F$ is full rank is in principle guaranteed as long as at least $N$ of the $M$ measurement points $\vc{q}_j$ are different. However, as will be shown shortly, if the sample distribution is not uniform the matrix can have an \emph{effective rank} much smaller than $N$, making the linear system highly ill-conditioned. 

The robustness to noise of relation \eqref{eq:Finv} is best investigated with a singular value decomposition of $F$, or equivalently by studying the eigenvalues and eigenvectors of $F^{\dagger} F$. This problem was investigated in depth in the 1960s \cite{Slepian1961,Landau1967,DiFrancia1969}, especially for the problem of counting and characterizing the degrees of freedom transmitted in an optical system. 

Singular value decomposition (SVD) entails factorizing $F$ as
\begin{equation}
 F = USV,
\end{equation}
where $U$ and $V$ are $M \times M$ and $N \times N$ unitary matrices, and $S$ is a non-negative $M \times N$ matrix whose only non-zero entries, the singular values, are along the main diagonal. Upon substitution in \eqref{eq:xfy}, the system becomes diagonal:
\begin{equation}
 y' = S x',
\end{equation}
where $y' = U^{\dagger} y$ and $x' = Vx$ can be seen as a change of basis for $y$ and $x$. A zero singular value in $S$ means that the corresponding entry in $x'$ is not constrained by the measured data $y$. In practice, unconstrained degrees of freedom are those for which singular values are smaller than a threshold typically defined by uncertainty in measurements. In this sense, the number of singular values above the threshold plays the role of an effective rank for $F$, and the rows of $V$ corresponding to singular values lower than the threshold represent the effective null space of $F$.

\begin{figure*}%
\centering
\subfloat[]{%
\label{fig:disk}%
\includegraphics[width=.35\textwidth]{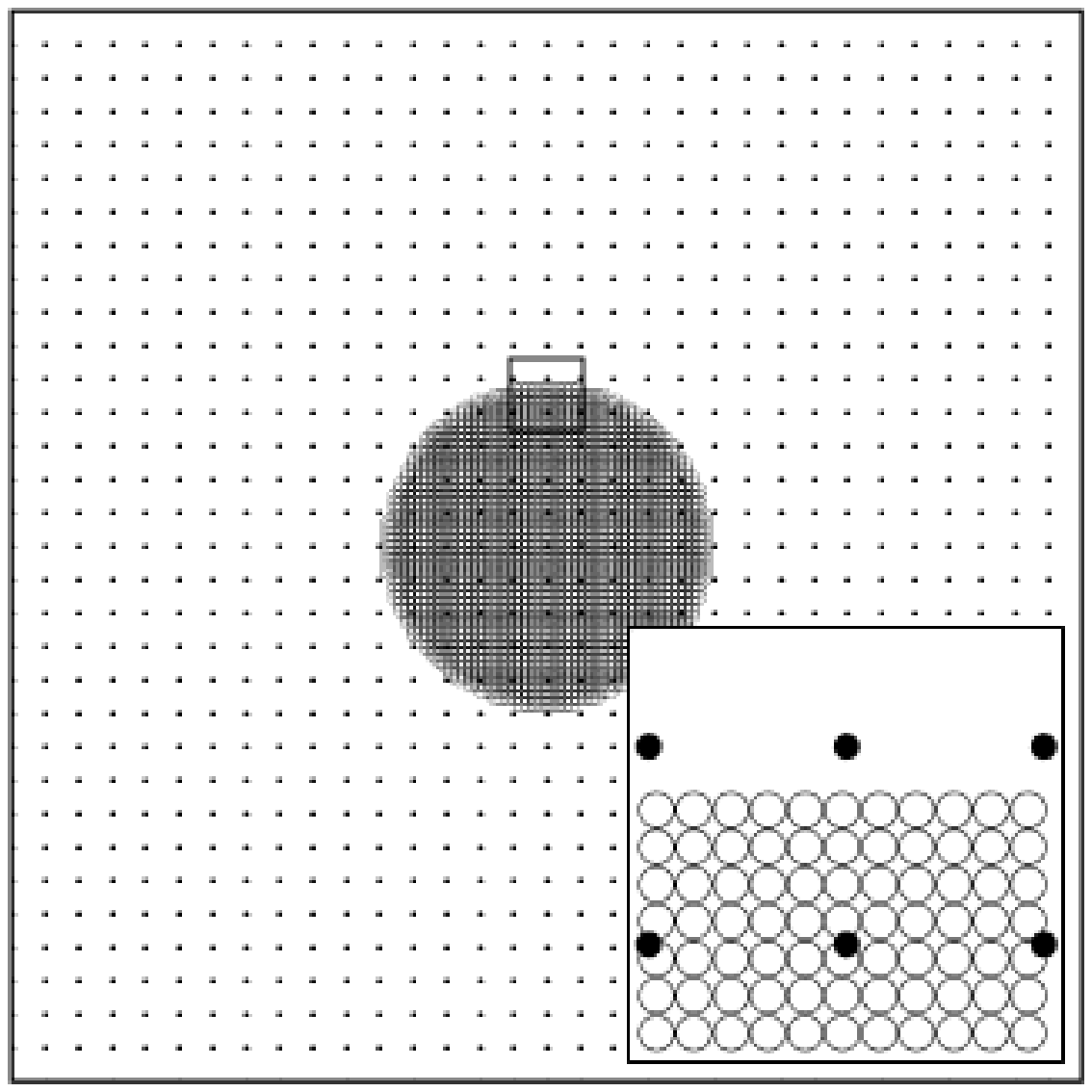}%
\vspace{-1cm}
}%
\hspace{8pt}%
\subfloat[]{%
\label{fig:disk_svd}%
\includegraphics[width=.35\textwidth]{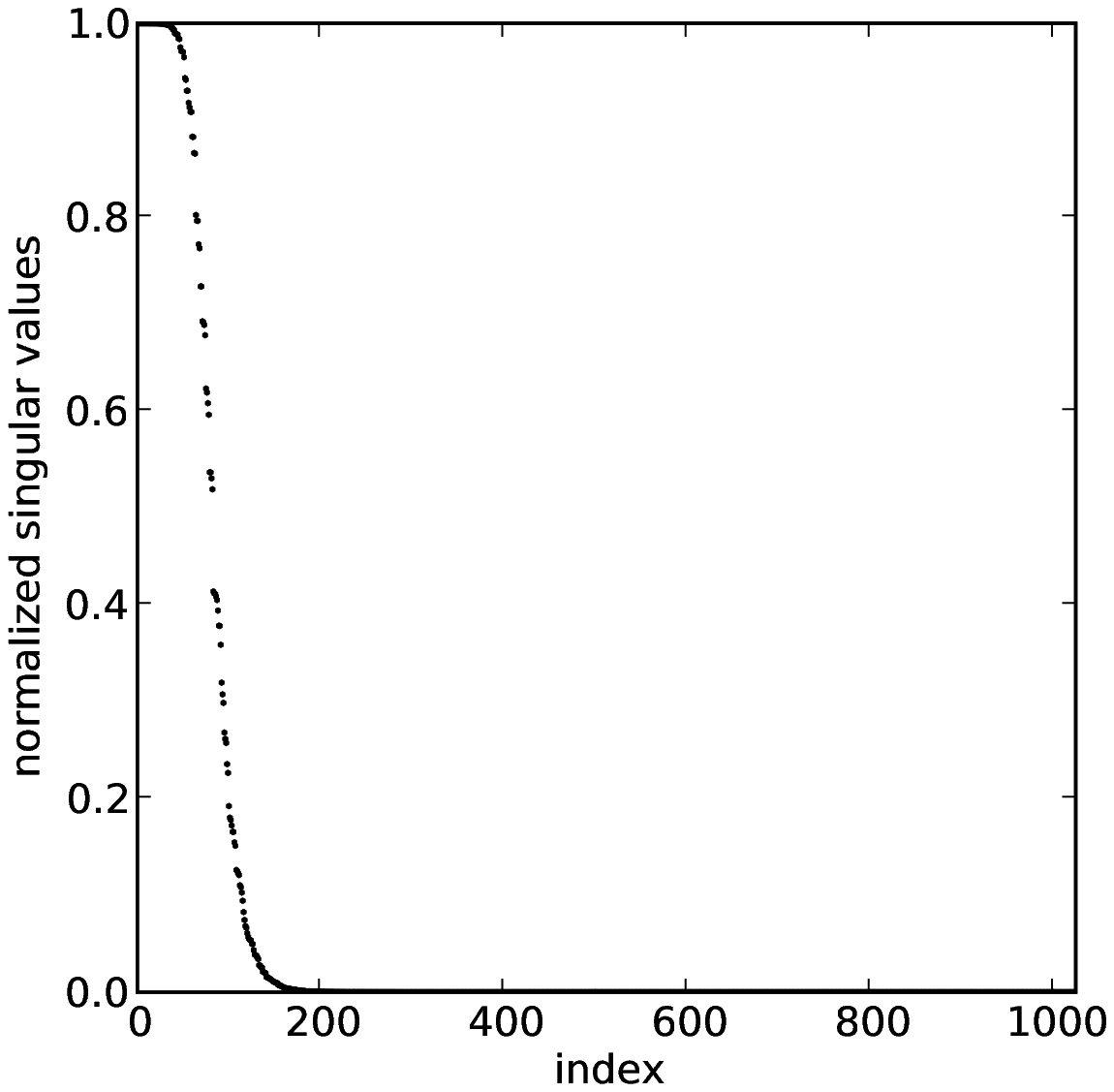}%
\vspace{-1cm}
}\\
\vspace{-.4cm}
\subfloat[]{%
\label{fig:random}%
\includegraphics[width=.35\textwidth]{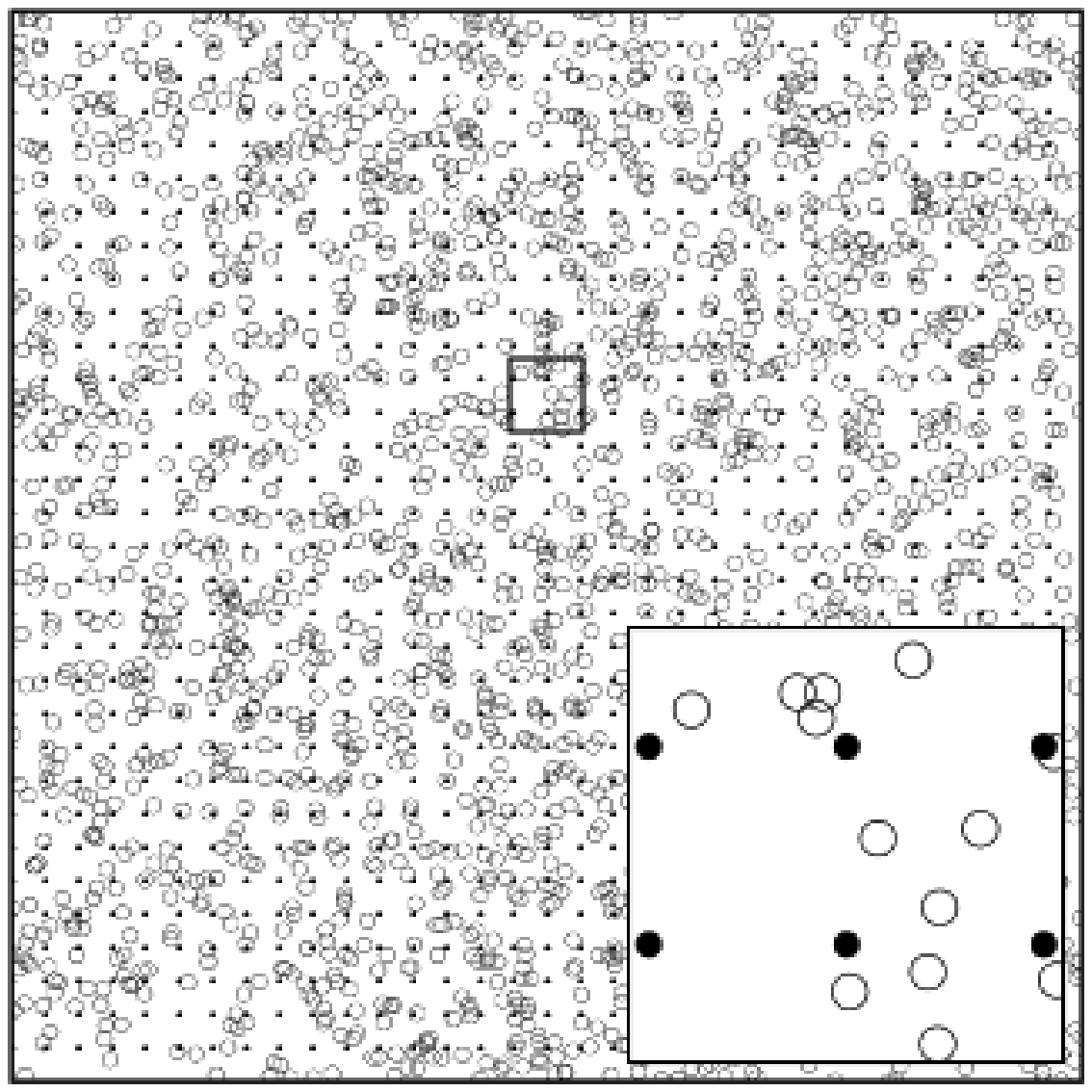}%
}%
\hspace{8pt}%
\subfloat[]{%
\label{fig:random_svd}%
\includegraphics[width=.35\textwidth]{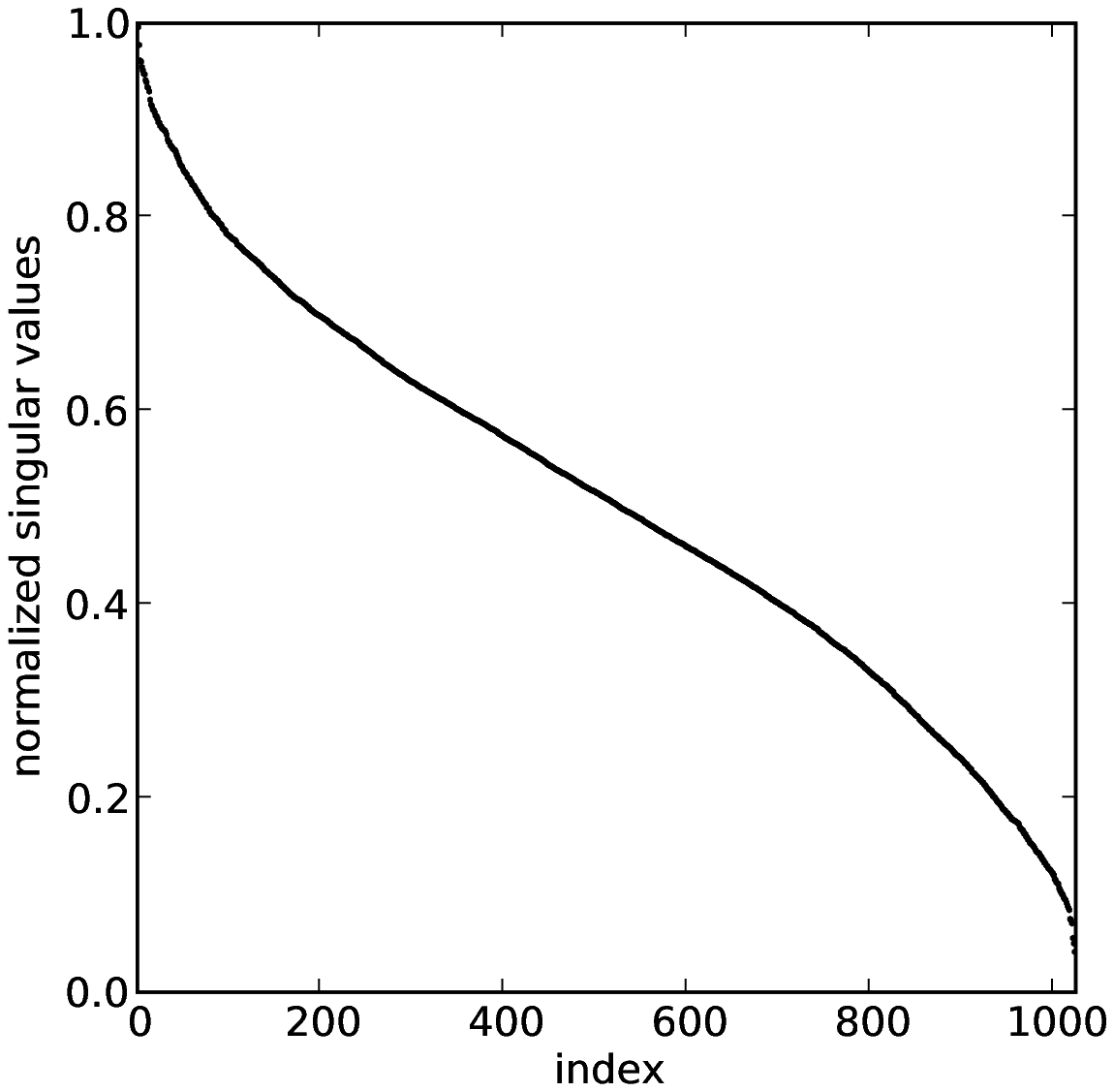}%
}\\
\vspace{-.4cm}
\subfloat[]{%
\label{fig:Ewald}%
\includegraphics[width=.35\textwidth]{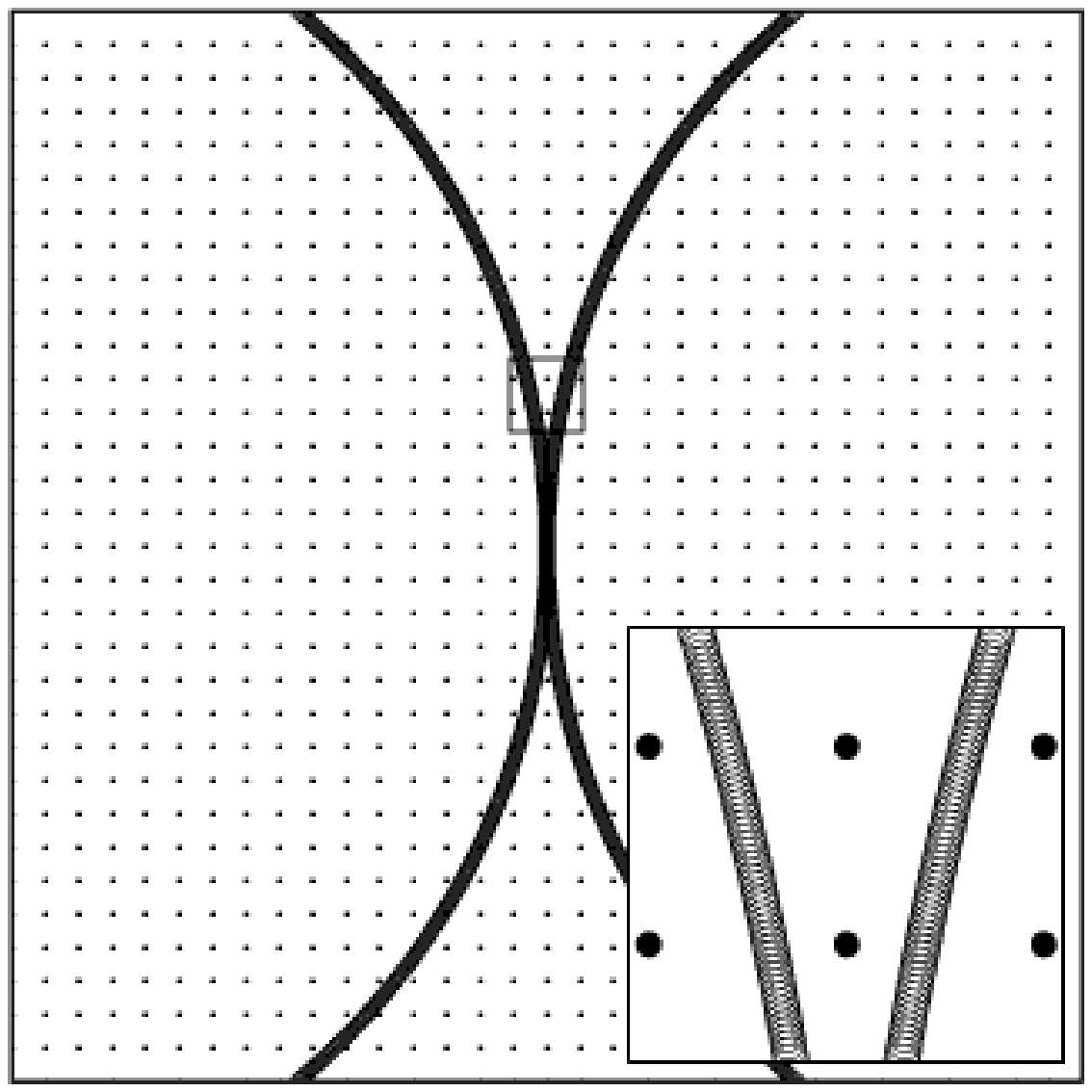}%
}%
\hspace{8pt}%
\subfloat[]{%
\label{fig:Ewald_svd}%
\includegraphics[width=.35\textwidth]{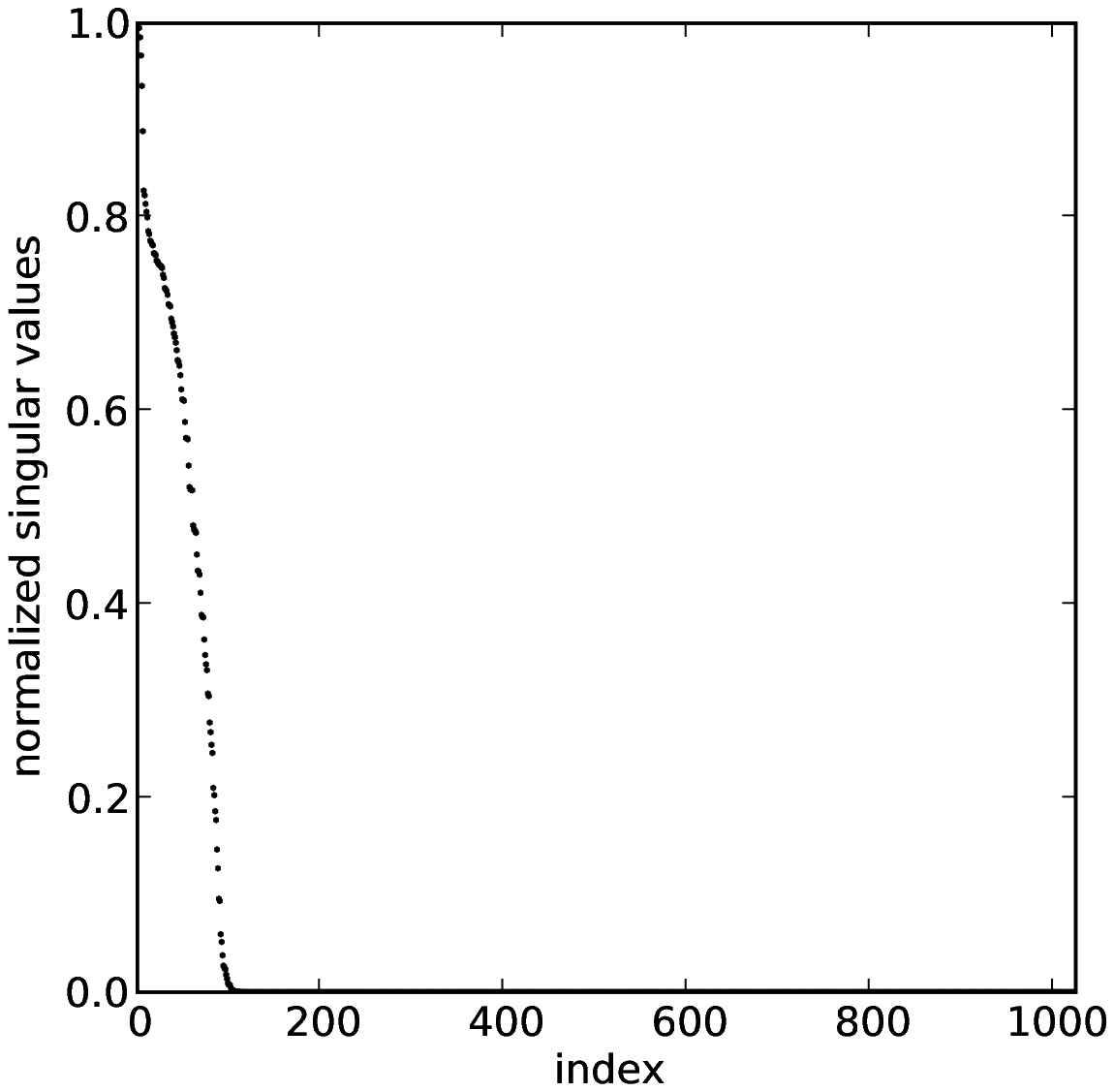}%
}
\caption{Examples of non-uniform samplings and the resulting singular value decompositions. In all cases, the regular $32\times32$ grid (dots) is at the Shannon sampling density required for the real space support (not shown) to be completely determined. (a) Sampling limited to a disc. (c) Random sampling positions. (e) Two-dimensional equivalent of the Ewald sampling. The sampling points are represented by open circles.}
\label{fig:svd}
\end{figure*}

We illustrate the behavior of the singular value decomposition of $F$ with a simple two-dimensional system. The two-dimensional function $A$ has a square support and is reconstructed at a resolution such that its extent is $32 \times 32$ pixels, so that $N = 1024$. Three distributions of measurement points were considered. In all cases, the number of measured samples was $M=2N$, so that the problem is in principle well-posed. Figure \ref{fig:svd}(a) reproduces the classical problem of imaging with a finite size pupil. All measurement points $\vc{q}_j$ are contained within a disc. The singular values for this system exhibit a relatively sharp transition around an index called the ``Shannon number'' \cite{DiFrancia1969}. The second case, where the positions $\vc{q}_j$ have been assigned randomly, shows a very different singular value distribution, indicating that the interpolation is well-conditioned. The last case, shown in Fig. \ref{fig:svd}(e) is a two-dimensional equivalent of an Ewald sphere measurement with total scattering angle of $100^{\circ}$ and including the Friedel pairs since $A$ is real-valued. The corresponding plot of the singular values shows clearly that out of the $1024$ degrees of freedom in $A$, at most $100$ can be determined reliably in realistic noise conditions. 

In the first and third cases, the measured domains are practically connected. For such cases the Shannon number, like Weyl's formula \cite{Weyl1911}, typically scales like the volume (or area) of the measured set, and is independent of the sampling density within the set, as long as it is higher than the Nyquist density.

In ``ankylography'', the area of the measured Ewald sphere portion in Fourier space is 
\begin{equation}
 A_{\text{Ewald}} \approx 2 \pi k^2 \left( 1 - \cos \theta_{\text{max}} \right),
\end{equation}
where $\theta_{\text{max}}$ is the maximum scattering angle recorded. To obtain a good proportionality factor, this area has to be expressed in units of typical ``speckle size'' $s = \pi/w$, where $w$ is the extent of the object's support, assumed to be the same in all directions. Up to a factor of order unity, the number of constrained degrees of freedom is thus
\begin{equation}
 N_S \approx 8 \pi \frac{w^2}{\lambda^2} \left(1 - \cos \theta_{\text{max}}\right),
\label{eq:Snum}
\end{equation}
where $\lambda = 2 \pi/ k$ is the radiation wavelength. An approximation for small scattering angles and upper bound for \eqref{eq:Snum} is
\begin{equation}
 N_S \approx 4  \pi \frac{w^2}{\Delta x^2},
\label{eq:NS_Ewald}
\end{equation}
where $\Delta x$ is the pixel size in the reconstruction of $\rho$. Not surprisingly, the number of degrees of freedom fixed by one Ewald sphere measurement scales like the area of a section of the reconstruction rather than its volume. In other words, only a small fraction (roughly $4 \pi \Delta x/w$) of the degrees of freedom are constrained by the measurement. Seeing $N_S$ as an effective rank for $F$ suggests that the high sampling rate imposed by the condition $M>N$ is effectively meaningless as soon as noise is introduced. To illustrate this point, Fig. \ref{fig:svd2}(a) shows a case with $M=128$. The resulting singular values, plotted in Fig.\ref{fig:svd2}(b) (full line) agree very well with the case $M=2048$ (points), and the difference between them becomes relevant only when exceptionally high signal-to-noise ratio can be achieved. On the same figure are also plotted the singular values for $M=64$ (dotted line) and $M=96$ (dashed line).

\begin{figure}[!b]%
\centering
\subfloat[]{%
\includegraphics[width=.4\textwidth]{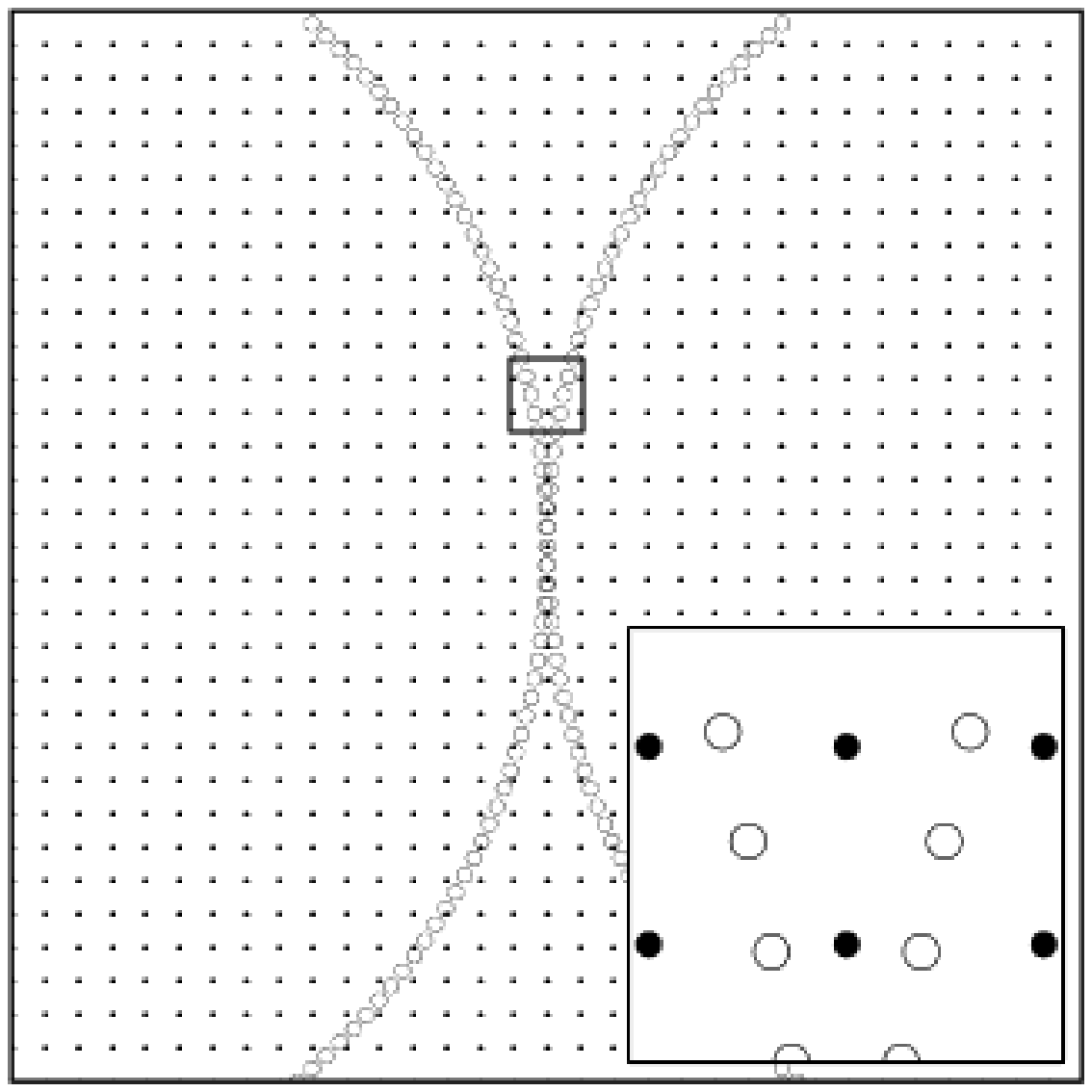}%
}%
\hspace{8pt}%
\subfloat[]{%
\includegraphics[width=.4\textwidth]{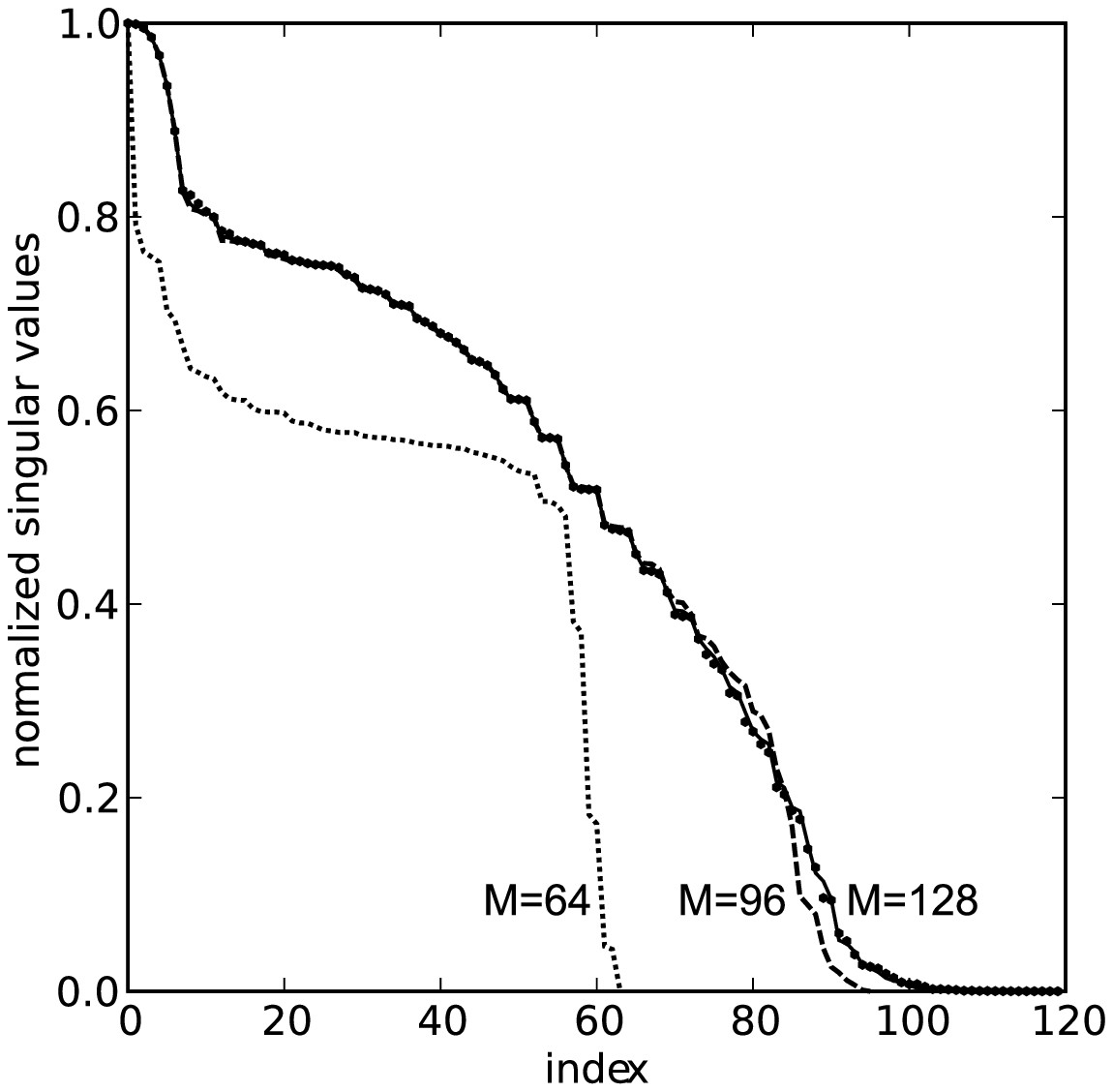}%
}
\caption{Two-dimensional Ewald sampling with lower density than the full-rank condition (see main text).}
\label{fig:svd2}
\end{figure}

The analysis developed above focused on the problem of recovering the autocorrelation from the intensity measurement because the problem is linear and does not involve the additional complications of the phase recovery problem. Obtaining the full autocorrelation as a first step is a sufficient condition for the phase problem to have a unique solution \cite{Bates1982}, but it certainly is not a necessary condition, since the number of unknown in $\rho$ is at least $4$ times smaller than the unknowns in $A$. The analysis is still useful even if, for instance, one assumes that the phases are known on the Ewald sphere, since the sampling conditions are the same when $I$ is replaced with $\tilde \rho |_{\text{Ewald}}$ and $A$ is replaced with $\rho$. 

With regard to the manuscript \Raines, a few conclusions are drawn from this section. First, the theoretical tools to discuss the extrapolation problem are already available. These tools are much more accurate and efficient than the rather vague constraint set heuristic used by the authors. Of course much more can be done within this formalism than the simple examples shown above. For instance, one could correlate directly the measurement noise with the threshold to compute a more accurate effective rank for $F$. In addition, uniqueness issues can be probed using the singular vectors obtained in the SVD calculation. These vectors form a basis on which an image can be decomposed to characterize the spatial distribution of each degree of freedom. This approach is similar to the weakly constrained mode analysis developed to describe loss of information caused by the central missing data in diffraction microscopy \cite{Thibault2006}.

A second conclusion is that ankylography, as a general method, is doomed to failure most of the time. As illustrated by equation \eqref{eq:NS_Ewald}, the data effectively provides only a small portion of the constraints that are required for the problem to be well-posed. As a consequence, in practical situations the ``oversampling degree'' condition $O_d>1$ stated in \Raines~is unnecessary, and can be replaced with $M > N_S$. The problem then becomes gathering sufficient additional information to ensure the uniqueness of the solution. Whether enough is known about the sample in addition of the data is probably the most critical and most difficult question to answer. The next section addresses, among other things, these necessary complementary constraints. 







\section{Reconstruction algorithm}

Retrieval of the three-dimensional density is without a doubt a difficult task, since it involves searching for a unique solution in a large phase space by ensuring that all known constraints are satisfied at once. In traditional diffraction microscopy, two constraints -- data in Fourier space and support in real space -- typically suffice to determine the solution uniquely. We have shown above that the problem in ``ankylography'' needs stronger constraints. Apart from the support, which can be refined as the reconstruction progresses, positivity of the reconstructed density is probably the most important constraint for the success of the reconstruction. Positivity can always be assumed unless the sample is made of a mixture of chemical species with very different indices of refractions (more precisely, for which the ratio $\delta/\beta$ is not the same). In the case of small clusters of atoms, it could be possible to impose an atomicity constraint, which can be very powerful if the resolution is sufficient. In Fourier space, another known constraint is the radial decay of the power spectrum, which helps controlling unconstrained degrees of freedom least correlated with the data, \textit{i.e.} located away from the Ewald sphere region.

Among other restraints discussed in \Raines~are ``continuity'' and ``uniformity'' operations. The ``continuity'' operation entails multiplying $\tilde \rho$ with a Gaussian function. That it is a perturbation that favors the convergence of the algorithm, as claimed by the authors, may have to be demonstrated. Instead, the main contribution of this operation appears to be a control on the radial decay of the amplitudes. Essentially a smoothing operation, the operation is rather ill-named, as it pertains little to the mathematical definition of continuity.  The ``uniformity'' operation would require more clarification from the authors. Uniformity outside the support region appears to be guaranteed by the support projection alone, which consists in setting all pixels outside the support to $0$. 

Compared to other commonly used methods, the structure of the iterative reconstruction algorithm presented in \Raines~is difficult to interpret. It is defined as multiple nested loops where various operations are applied in turn. Unlike most commonly used iterative algorithms \cite{Fienup1982,Elser2003a,Luke2005}, many of the operations are not projections onto constraint sets. The projections formalism, also widely used in convex optimization problems \cite{Youla1982,Bauschke1996}, allows a precise characterization of the reconstruction dynamics, and helps interpreting the outcome of a reconstruction. Using projections, an iterative algorithm is not merely a ``recipe that happens to work well''. One of the requirements for an operation to be a projection is idempotency: applying a projection twice is the same as applying it once. It is this characteristic that is missing in the ``continuity'' and ``uniformity'' operations. As a result, the frequency at which they are applied affects the final solution. If, for instance, the Gaussian multiplication is applied more often, its effect will accumulate and the high spatial frequency amplitudes are expected to be smaller. Questions such as: how often should this operation be applied? how is ``convergence'' defined? How is ``solution'' defined? remain widely unanswered. The seemingly \textit{ad hoc} succession of operations makes very difficult the identification of the algorithm's strengths and weaknesses.


\section{A flawed experimental demonstration}

We complete this report with a discussion on the reconstruction of experimental data presented in \Raines. A diffraction pattern was measured using an extreme ultraviolet (EUV) laser source ($\lambda = 47 \text{ nm}$). The sample was a figure etched through a $100 \text{ nm}$ silicon nitride membrane. At this wavelength, the transmission of the membrane is negligible.
The diffraction pattern was collected with a CCD placed near the sample to collect scattering out to a large angle. In \Raines, Figure 4 shows an isosurface rendering of the three-dimensional reconstruction obtained from this single diffraction pattern.

It may seem odd at first that the reconstructed density is really that of a hole in a membrane. What is the physical origin of the front and back faces of the reconstruction if the sample was in reality an empty region sandwiched between nothing? That such a measurement is in principle possible is a consequence of a three-dimensional version of Babinet principle. It suffices to express the three-dimensional density of the etched membrane as the difference between a uniform slab and the density of the material that would form the figure in the hollow mask: $\rho_{\text{mask}} = \rho_{\text{slab}} - \rho_{\text{figure}}$. After taking the Fourier transform on both sides of this equation, one observes that the slab term has its power concentrated on a single line perpendicular to the slab surface (a truncation rod), which intersects the Ewald sphere only at the origin. Therefore at the exception of the very center of the diffraction pattern, the scattering from the mask is exactly equal to the scattering from an isolated absorbing figure having the same shape as the hole.

\begin{figure}%
\centering
\subfloat[]{%
\label{fig:disk}%
\includegraphics[width=.4\textwidth]{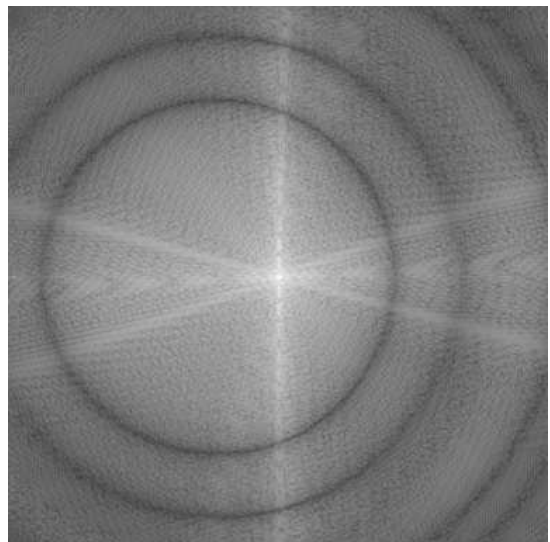}%
}%
\hspace{8pt}%
\subfloat[]{%
\label{fig:disk_svd}%
\includegraphics[width=.4\textwidth]{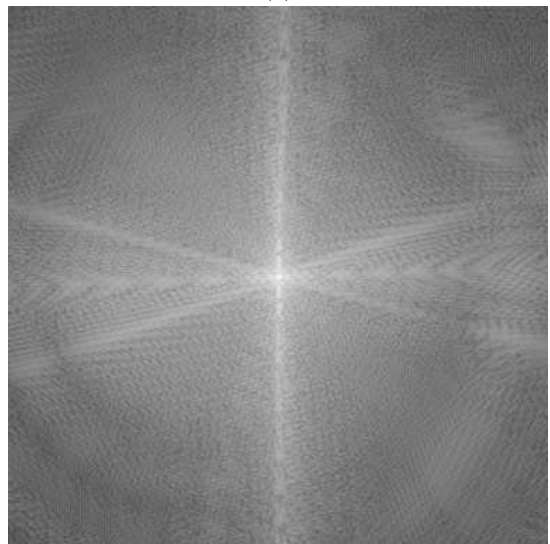}%
}
\caption{Multislice simulation of the diffraction from a mask etched in a membrane tilted by $5^{\circ}$ relative to the incoming plane wave. (a) Weakly scattering object. The membrane transmits $100\%$ of the wave and causes a phase shift of about $0.02\pi$. (b) Strong object. The transmission of the membrane is $.03\%$.}
\label{fig:membrane}
\end{figure}

This explanation however is not applicable to the reconstruction presented in \Raines, because the absorption of the membrane is very large, and thus the Born approximation is not valid. Consequently, apart from eventual dynamical scattering along the edges of the hole, no information on the thickness of the membrane can be transmitted to the detector. Figure \ref{fig:membrane} shows diffraction patterns obtained by multislice simulations of the same mask but with different scattering strengths. The weakly scattering membrane in Fig \ref{fig:membrane}(a) yields characteristic dark rings that are readily interpreted as the intersection of the Ewald sphere with the zeros of the sinc function from the Fourier transform of the slab density. In the present case the rings are off-center because the membrane was tilted by an angle of $5^{\circ}$ to reproduce the experimental conditions of \Raines. The rings are absent in Fig. \ref{fig:membrane}(b), which shows the strongly absorbing case. 

Strong absorption is not compatible with the single-scattering view of Ewald sphere sampling. The reconstruction presented in \Raines~is therefore not valid. That the overall tilt of the membrane could be reconstructed should not come as a surprise, as this information is recorded in the distortion of the diffraction pattern on the flat detector. It appears that the reconstruction presented in \Raines~is in reality a tilted 2D reconstruction, in which the depth of the flat sample is just an artifact -- an extrusion of the two-dimensional map. A good experimental proof of ankylography would have been to produce two weak objects placed a suitable distance apart in the propagation direction. 

\section{Conclusion} 

The main conclusions of this ``open review'' of Raines et al. \cite{Raines2009} are: (1) Although valid in principle, ankylography will probably succeed only in special cases where high-quality data and strong additional constraints are available; (2) The description and the properties of the iterative reconstruction method may not meet the standards of state-of-the-art algorithms in the field; (3) The experimental demonstration is not valid because the sample does not satisfy the Born approximation.

\section{Acknowledgements}

Many thanks to Veit Elser for useful discussions and advice.

\newpage
\normalsize

\onecolumngrid

\appendix
\section{Open letter in reply to the authors' reply}

This letter was originally made available at the address

\texttt{http://people.web.psi.ch/thibault/current/letter\_to\_John\_Miao.pdf}.

\vspace{1cm}
\hfill\parbox[t]{5.5cm}{Pierre Thibault\\
Department of physics\\
Technical University of Munich}

\noindent John Miao\\
Department of Physics and Astronomy\\
UCLA\\

\noindent Dear John,

\vspace{1cm}

Here is my short answer to the response you produced \cite{Miao2009response} about my commentary article \cite{Thibault2009comment} on ``ankylography'' \cite{Raines2009}. I thank you for taking the time of preparing this answer. It will certainly be useful to make my manuscript clearer in the next version. I was nevertheless disappointed to see that you dismissed all of my criticisms. I feel that many of my points were misunderstood and some even misrepresented.

\vspace{1cm}

You seemed to refuse to consider most of my theoretical derivation for two main reasons: (1) that phase retrieval can be possible even if the intensity measurement is partly aliased and (2) that additional real-space constraints are necessary. 

First, I want to say that I agree with these statements and that a rectification is in order. Regarding (1), you said ``Thibault implied the requirement of obtaining the autocorrelation function for phase retrieval of coherent diffraction intensities, which was the basis of his theoretical analysis'' (page 2). Still, I thought that I made clear that I was aware of this fact by saying that ``obtaining the full autocorrelation as a first step is a sufficient condition for the phase problem to have a unique solution, but it certainly is not a necessary condition'' (page 9).

As for (2), you wrote that my analysis is ``flawed'' because I have ``ignored the vital importance of the various physical constraints'' (page 7). Yet, I wrote that ``the data effectively provide only a small portion of the constraints that are required for the problem to be well-posed [...] The problem then becomes gathering sufficient additional information to ensure the uniqueness of the solution. Whether enough is known about the sample in addition of the data is probably the most critical and most difficult question to answer'' (page 9).

More important is the fact that these two statements do not affect the results of my analysis. The goal of the section about non-uniform sampling was to describe \textit{and quantify} how much, or how little, information on the object is provided by the intensity measurement. I obtained an expression for the rough number of degrees of freedom fixed by the data alone, and specified, as quoted above, that additional degrees of freedom would need to be restrained with real-spaced constraints.

I found your argument about the autocorrelation very puzzling. Even though it is true that some degree of aliasing still permits reconstructions, this is hardly relevant in the present discussion since we are free to choose the sampling of the intensity, and there is no good reason to pick a sampling density that is lower than the Nyquist criterion. All simulations in your paper certainly were far away from the lower end of the possible sampling density, as is the Figure 1 in your response. Note also (looking at Figure 1 below) that the situation of a lower sampling density can just as well be described in the formalism I described in my comment. 

\begin{figure}[!b]%
\centering
\subfloat[]{%
\label{fig:disk}%
\includegraphics[width=.45\textwidth]{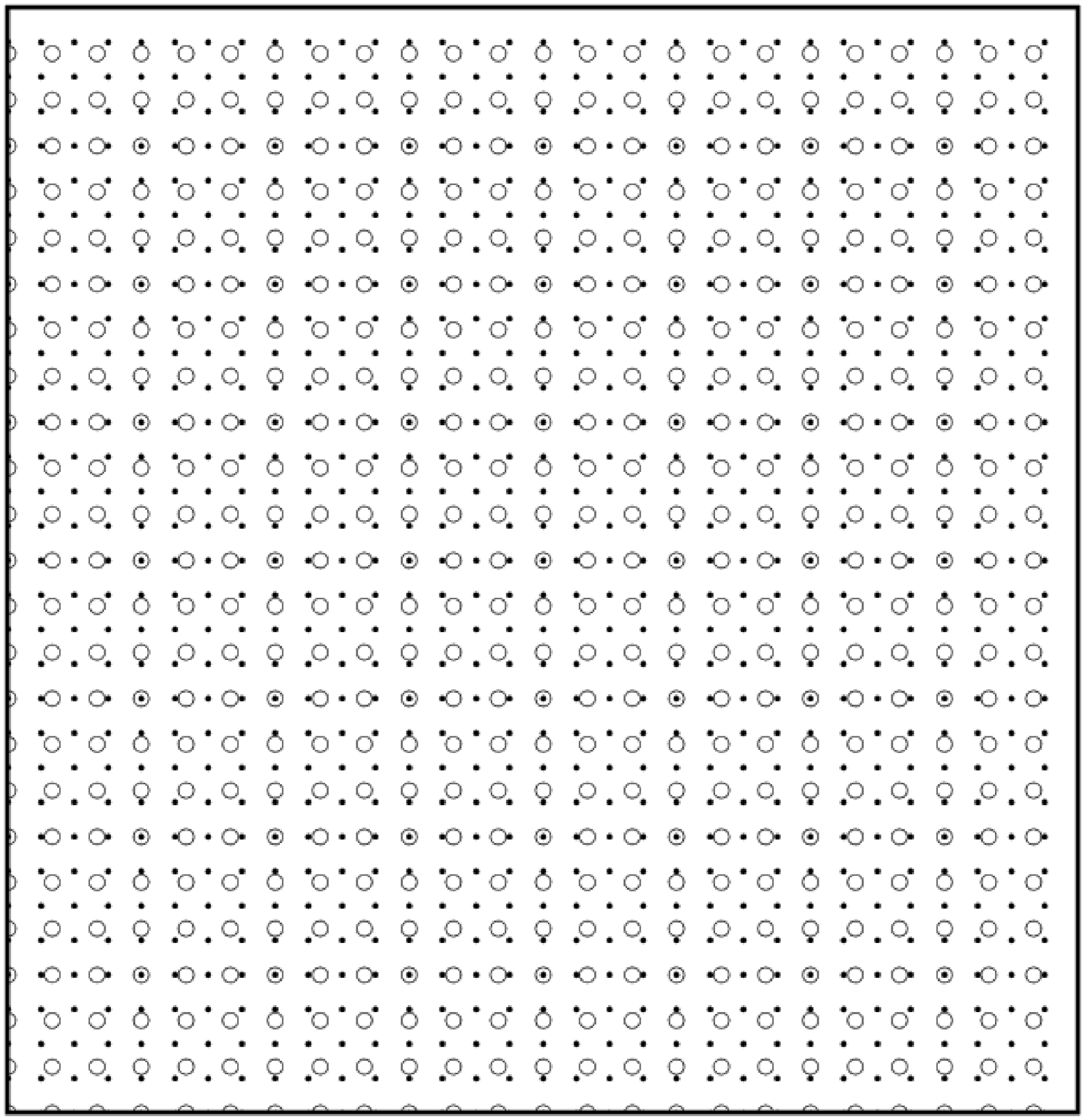}%
}%
\hspace{8pt}%
\subfloat[]{%
\label{fig:disk_svd}%
\includegraphics[width=.45\textwidth]{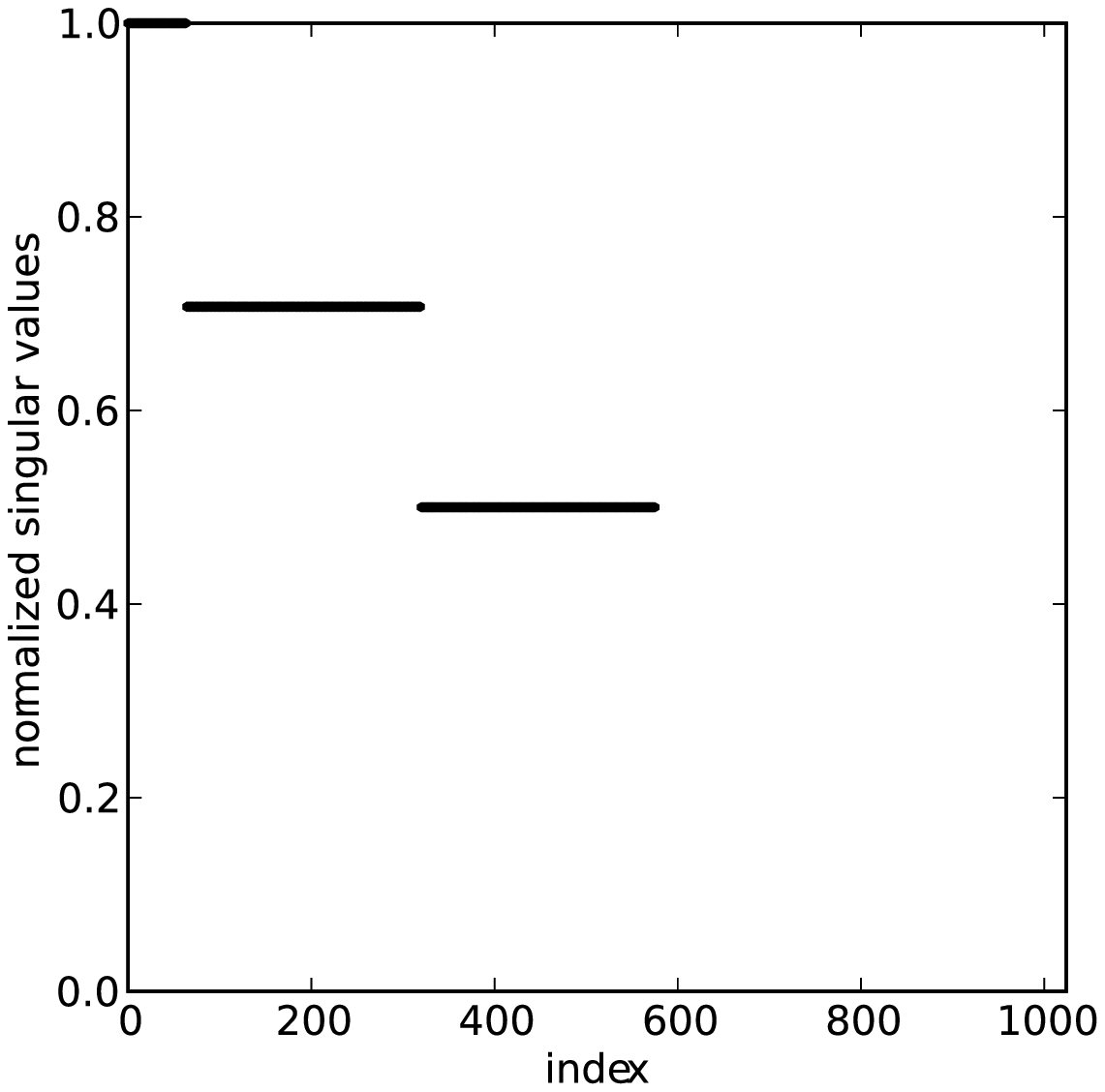}%
}
\caption{(a) Representation of a sampling density lower (circles, $24 \times 24$) than that required by Nyquist criterion (dots, $32 \times 32$). (b) Singular values resulting from the SVD calculation on this system. All singular values are large but the system has only $24^2 = 576$ fixed degrees of freedom. Taking into account the centrosymmetry of the autocorrelation, the actual number of independent degrees of freeedom fixed by the measurement is $288$, which should still make phase retrieval possible as the object has $16^2 = 256$ independent pixels ($\Omega = 288/256 > 1$ or equivalently in this situation $\sigma = 576/256 > 2$).}
\label{fig:svd2}
\end{figure}

This freedom of choosing the sampling density is a direct consequence of Shannon's theorem: if Nyquist criterion is satisfied, the signal is \textit{completely} determined by its samples. In your paper and in Figure 1 of your response, you chose a sampling substantially higher than the Nyquist criterion. Still, I want to emphasize that any sampling that is denser than Nyquist's criterion is redundant and contains no new information. Keeping this fact in mind, I hope that you now see how the Figure 1 in your response is completely equivalent to my Figure 1(e), the only difference being that I picked the coarsest sampling allowed by the Nyquist criterion. That's what I meant in the caption by ``the regular $32 \times 32$ grid (dots) is at the Shannon sampling density required for the real space support (not shown) to be completely determined.''. 

Looking carefully at your response, I sense that the main source of misunderstanding is about a quite fundamental question, namely how to count the constraints and the degrees of freedom in diffraction microscopy in general. It used to be thought that the oversampling ratio,
\begin{equation}
\sigma = \frac{\text{number of intensity measurements}}{\text{number of independent pixels in the support}},
\end{equation}
is the right way of measuring how overconstrained is a problem. This definition seems intuitive: for a given object size, the more individual intensity measurements, the more the system is constrained. But is it not completely correct.
As mentioned above, Shannon's theorem says that sampling a signal at a higher frequency than the Nyquist criterion adds no information. So it is not correct to say that an intensity measurement with $\sigma=20$ contains twice as much information as the measurement with $\sigma=10$ since, beyond Nyquist frequency, any additional sample depends linearly on the other samples. It turns out that the real number of constraints in the problem is not the number of intensity measurements, but rather the number of independent pixels in the autocorrelation support (normally $1/2$ the total number of pixels in the autocorrelation support because it is always centrosymmetric). This leads to the definition of a more meaningful ratio:
\begin{equation}
\Omega = \frac{\text{number of independent pixels in the autocorrelation support}}{\text{number of independent pixels in the support}}.
\label{eq:Omega}
\end{equation}
The requirement for reconstructibility is $\Omega > 1$. I invite you to take a look at the recent paper by Elser and Millane \cite{Elser2008}, where the properties of this ``constraint ratio'' are explored in details. Note that $\Omega$ becomes equal to $\sigma/2$ in the case where the autocorrelation starts to ``wrap around'' because of a lower sampling rate. This situation is illustrated in Fig. 1 above. In the case depicted, all singular values of the matrix $F$ are large enough for the system to be invertible. Still the rank of $F$, and thus the number of fixed degrees of freedom, is limited by the number of intensity measurements. At this lower end of the sampling density, $\sigma$ becomes a good measure of how difficult is a reconstruction, as you, David Sayre and Henry Chapman have shown a little while ago \cite{Miao1998}. Yet, in pretty much all practical cases, the sampling of the intensity satisfies the Nyquist criterion. Sampling on a finer grid has clear experimental benefits, as a way of increasing the dynamic range and reducing the effect of the detector point spread function. But it does not make solving the phase problem fundamentally easier.

To summarize: the maximum number of degrees of freedom that can be fixed by the data is the number of independent pixels in the autocorrelation. The point of my analysis was to find what proportion of these degrees of freedom was effectively fixed by a single Ewald sphere measurement. What I wanted to show is 
\begin{enumerate}
\item that there is a formal way to describe why ankylography is possible \textit{in principle}, since I was unsatisfied with the set-theoretic argument used in your paper. Saying that ``the intensity points encode information from all possible orientations of a 3D object'' says very little on whether this information is sufficient, or if it can be extracted.
\item that \textit{in practice}, the problem is highly ill-conditioned, so that the condition $O_d > 1$ is not relevant to realistic datasets, and that the Shannon number is a better measure than the number of intensity measurements to estimate the number of degrees of freedom fixed by the data.
\end{enumerate}

\vspace{1cm}

I will not dwell too long on my criticism of your algorithm. I just want to point out that the ``projection formalism'' to which I was referring is much more general than you implied. I was of course not advocating the use of ER instead of HIO. What I meant is that all of the most widely used algorithms (error reduction, hybrid input output, difference map and relaxed averaged alternating reflections) can be express in this formalism, which is very convenient to understand their dynamics \cite{Marchesini2007}. 

\vspace{1cm}

Your response regarding the experimental demonstration is certainly the part with which I disagree the most. I must not have been clear enough.

First, I should mention that the second paragraph about the ``3D Babinet'' principle was little more than a pedagogical step, a mere preparation for my argument. I wanted first to answer an eventual question about the physical significance of the recovered 3D density. In a very simple-minded approach, one could ask: ``this etched figure is just an empty hole, how can you reconstruct the density of a hole, and if you do, what is its meaning? It certainly is not an electron density.'' To this hypothetical question, I answered that \textit{if the Born approximation is valid}, one can see it as the negative of the slab density, since a uniform slab affects only the very center of the diffraction pattern.  Calling the density subtraction a ``3D Babinet'' principle may have been misleading, but it does follow the same reasoning as in the 2D case. Now, I want to reiterate that the only goal of this paragraph was to establish that \textit{in the Born approximation} this interpretation holds, and that doing such a reconstruction is meaningful, even quantitatively.

I certainly ``realized that taking the Fourier transform on both sides requires [...] the Born approximation'' (page 9), as implied by the beginning of the next paragraph, where I stated that ``this explanation however is not applicable to the reconstruction presented in R09, because the absorption of the membrane is very large, and thus the Born approximation is not valid.'' This is the essence of my argument, which I would certainly not call ``self-contradictory''. By mapping the measured intensities on the Ewald sphere, ankylography relies on the Born approximation [this is the condition for equation (4) in my comment to be valid]. But a strongly absorbing sample does not satisfy Born approximation. Therefore it is wrong to interpret the measured intensities as a spherical section of the squared 3D Fourier transform of the density. In your reconstruction, the 3D array in which the measured intensities are distributed has no physical interpretation: it certainly is not the Fourier transform of the 3D electron density (or refractive index map) of the slab since this requires the Born approximation to be valid.

I don't know why you so hastily dismissed my numerical simulations as invalid. These are multislice simulations in which I did not do any assumption about a ``Babinet'' principle (obviously it would have been wrong, as you pointed out). I don't know exactly on what is based your judgment that Fig. 3(b) ``doesn’t look like a Fraunhofer diffraction pattern at all'', since that is exactly what it is: the free-space propagation to infinity of the exit wave generated by a multislice simulation of a wave passing through an aperture in a slab with total transmition of $3 \times 10^{-4}$. The slab was even tilted by $5^{\circ}$. The slow modulation of the intensity, reminiscent of the rings in Fig. 3a probably is the effect of the dynamical scattering of the edge (such as the point $P_3$ in your figure) which the multislice simulation does take into account. The main point of Figure 3 was to show the dark rings in Fig. 3(a), which are the signature of the validity of Born approximation which (let me repeat it again) is a prerequisite for ankylography to work.

You say that ``it can be easily shown by using the diffraction theory that, when the edge effect is negligible, the measured intensities are proportional to the square of the Fourier transform of the 3D sample on the Ewald sphere'' (page 10).
I don't think this is true and I would be glad to see your proof of that. What I contend instead is that the valid way of treating an absorbing mask is by assuming that it is a two-dimensional object -- it may be a tilted planar object, but still only planar. I can provide a detailed derivation if you want. This is what led me to conclude that apart from the tilt, which can be extracted from the data, the rest is only an artificial extrusion of a 2D amplitude. I am also ready to debate your affirmation that the ``structure defects would be invisible in the 2D reconstructed image'' (page 11), and would certainly be happy to look into your diffraction data to prove my point.

\vspace{1cm}

As I said above, I am sorry that you rejected all of my criticisms, and I hope to be able to discuss further any of them with you soon. I will also check the validity of my claims through the peer review process by submitting soon an improved version of my arXiv comment for publication. 

\vspace{1cm}

Best regards,

\vspace{1cm}

Pierre

\vspace{1cm}
\twocolumngrid

\bibliographystyle{apsrev}

\end{document}